# Exploring Generative AI Policies in Higher Education: A Comparative Perspective from China, Japan, Mongolia, and the USA

Qin Xie, University of Minnesota, USA

Ming Li, Osaka University, Japan

Ariunaa Enkhtur, Osaka University, Japan

## Abstract

This study conducts a comparative analysis of national policies on Generative AI across four countries: China, Japan, Mongolia, and the USA. Employing the Qualitative Comparative Analysis (QCA) method, it examines the responses of these nations to Generative AI in higher education settings, scrutinizing the diversity in their approaches within this group. While all four countries exhibit a positive attitude toward Generative AI in higher education, Japan and the USA prioritize a human-centered approach and provide direct guidance in teaching and learning. In contrast, China and Mongolia prioritize national security concerns, with their guidelines focusing more on the societal level rather than being specifically tailored to education. Additionally, despite all four countries emphasizing diversity, equity, and inclusion, they consistently fail to clearly discuss or implement measures to address the digital divide. By offering a comprehensive comparative analysis of attitudes and policies regarding Generative AI in higher education across these countries, this study enriches existing literature and provides policymakers with a global perspective, ensuring that policies in this domain promote inclusion rather than exclusion.

## Introduction

Generative Artificial Intelligence (AI) technologies like ChatGPT, Bing, and Co-Pilot have rapidly evolved into influential tools capable of generating human-like outputs (Bandi et al., 2023; Chan, 2023). Since OpenAI introduced ChatGPT in late November 2022, Generative AI has not only become a public sensation but also a transformative force across various sectors such as healthcare, finance, and entertainment, notably impacting higher education (Firaina & Sulisworo, 2023).

The debate over Generative AI's role in higher education is intensifying. Research shows that tools like ChatGPT can support educators in developing teaching materials, analyzing student data, detecting learning patterns, and enhancing teaching strategies (Kooli, 2023; Latif et al., 2023). Additionally, their utility in advancing accessibility and inclusivity for students with disabilities is recognized (Kasneci et al., 2023). Despite these benefits, concerns about the application of Generative AI in academic settings are increasing due to risks of propagating problematic content, biases, and harmful assumptions, which could adversely affect all stakeholders in education (Li et al., 2023; Li et al., 2024). Consequently, educational institutions worldwide have adopted divergent stances on Generative AI, ranging from outright bans to enthusiastic adoption (Holmes & Miao, 2023).

At the international level, UNESCO has taken a leading role in regulating AI in education. In 2023, it issued the "Guidance for Generative AI in Education and Research," designed to assist countries in crafting both long-term and immediate policies, ensuring that Generative AI adheres to a human-centered approach. This follows its earlier "AI and Education: Guidance for Policymakers" published in 2021. The European Union has also been proactive, releasing resources like the "Ethical Guidelines on the Use of AI and Data in Teaching and Learning for Educators" and the "Executive Summary of the Final Commission Expert Group on Artificial Intelligence and Data in Education and Training" in 2022, aimed at promoting ethical AI use and dispelling common misconceptions.

Nationally, UNESCO's 2023 review revealed that only about 20 countries have clearly defined AI ethics regulations as part of their AI strategies, touching on educational implications. Despite several international bodies and nations developing or proposing regulations to manage Generative AI, its significant reliance on vast amounts of data and computing power means that predominantly large tech companies and select nations currently dominate this space (Holmes & Miao, 2023). This raises the necessity for a timely examination of how countries from the Global North and South adapt to Generative AI in educational contexts.

## Research Questions

This study explores national strategies and perspectives on Generative AI in higher education across four countries: China, Japan, Mongolia, and the USA. The research aims to address the following questions:

1. How are China, Japan, Mongolia, and the USA responding to Generative AI in higher education?
2. What are the differences in their approaches to Generative AI in higher education?

## Definition

### Generative AI

As defined by UNESCO in 2023, "Generative AI" refers to "an artificial intelligence (AI) technology that automatically generates content in response to prompts written in natural-language conversational interfaces” (p.8). This includes the production of new content such as texts written in natural language, images, videos, music, and software code. Our study adheres to UNESCO’s definition of Generative AI.

### Generative AI Policies

Generative AI policies encompass the rules, guidelines, and frameworks that regulate the development, deployment, and utilization of Generative AI technologies. These policies operate at multiple levels, including international, national, and industry-specific standards. In our study, we specifically consider Generative AI policies as those established by national authorities, which include both mandatory regulations and non-mandatory recommendations, particularly in relation to higher education.

## Methodology

This study employs the Qualitative Comparative Analysis (QCA) method to examine the policies on Generative AI across four countries: China, Japan, Mongolia, and the USA. QCA is a case-sensitive, set-theoretic method that excels in analyzing complex phenomena within specific cases (Thomann, 2020).

We have selected these countries as a representative sample due to their distinct approaches and contexts in the use of educational technology. China is pushing forward with the integration of large-scale online education and AI under national directives, adopting a unique approach. Japan implements a digital transformation policy to revitalize its education and economy. Mongolia views AI and digital technology as an opportunity to enhance educational access and mitigate geographical challenges. Meanwhile, The USA boasts a highly developed ed-tech market with advanced applications of Generative AI in education.

The primary aim of QCA is to identify significant patterns or phenomena that require explanation (Schneider, 2016; Thomann & Maggetti, 2020). This study specifically aims to explore how four countries respond to Generative AI in education. We focus on comparing the policies issued by the Ministries (Department) of Education or government authorities in each country.

**Data Collection**

To gather relevant information, the researchers employed a set of keywords including "Generative AI," "guideline," "policies," and "higher education" in both English and the local languages of Chinese, Japanese, and Mongolian. Searches were conducted on the official websites of the Ministries (Department) of Education in four countries. We successfully collected policies and guidelines from China, Japan, and the USA. Mongolia is currently in the process of developing official guidelines for the use of Generative AI in education. However, no regulation has been published on the Ministry website as of April 2024. Instead, we find a series of events and project launches incorporating "AI and digital technology" in education as well as other sectors as part of its development strategy (e.g. MONTSAME, 2024). Therefore, in this analysis, we include news and other information published on government websites.

**Data Analysis**

The analysis of the collected data was structured into two comparative steps. First, the contexts of the policies were compared, and in the second step, the focus shifted to the content of the policies. The researchers identified key thematic concepts such as "human-centered," "attitude," "teaching and learning," and "diversity, equity, and inclusion (DEI)," using both inductive and deductive reasoning to guide the analysis. The process is inherently iterative, involving continuous interaction between theoretical concepts and empirical evidence (Thomann & Maggetti, 2020).

## Findings

### Comparison of contexts

#### *Release time and issuing authorities*

The policy documents from China, Japan, and the United States were all released within a close timeframe, with the USA documents issued in May 2023, and those from China and Japan in July 2023 (Table 1). The issuers in Japan and the U.S. are similar, with both stemming from national education authorities. In contrast, China's guidelines were a collaborative effort, issued by the Ministry of Education alongside seven other departments. Mongolia has yet to address any policies on Generative AI. However, the Ministry of Digital Development and Communication (MDDC, n.d.) released documents in 2021 outlining plans to adopt AI as part of national goals aimed at transforming into an e-nation by 2050.

**Table 1 Basic information of four countries' policies/ guidelines**

| | Release Time | Release Documents | Release Authorities | Target Audiences | AI/ Generative AI | Educational Stage |
|---|---|---|---|---|---|---|
| **China** | July, 2023 | 生成式人工智能服务管理暂行办法 [Interim measures for the management of generative artificial intelligence services] | State Internet Information Office; National Development and Reform Commission PRC; Ministry of Education PRC; Ministry of Science and Technology of the PRC; Ministry of Industry and Information Technology of the PRC; Ministry of Public Security of the PRC; State Administration of | Generative artificial intelligence service provider, user | Generative AI | Did not clarify |
| | | | Radio and Television | | | |
| **Japan** | July, 2023 | 大学・高専における生成 AI の教学面の取扱いについて [About the Handling of Generative AI in Teaching at Universities and Technical Colleges] | Ministry of Education, Culture, Sports, Science and Technology, JAPAN | Governors, administrators, and faculty members in university and national institute of technology (KOSEN) | Generative AI | Higher education |
| **Mongolia** | September,<br><br>2021 | Vision-2050 Introduction to Mongolia's Long-Term Development Policy Document<br><br>"Alsyn haraa" [Strategy] in digital development and communications | Government of Mongolia<br><br>Ministry of Digital Development and Communications | Not clear | AI (not including Generative AI) | Mainly attribute to life-long education |
| **The USA** | May, 2023 | Artificial Intelligence and the Future of Teaching and Learning, Insights and Recommendations | Department of Education, Office of Educational Technology, US | Teachers, educational leaders, policy makers, researchers, and educational technology innovators and providers | Focus on broadly AI (including Generative AI), rather than a specific AI tool, service, or announcement | Specially focus on K-12 |

***Target audiences, technology focus, and educational stages***

In China, the documents primarily target service providers and users of Generative AI. While Mongolian documents target mostly users and learners of AI, the documents specify "using AI in life-long education" (Vision2050, n.d., p.62). Japan's guidelines are directed at governors, administrators, and faculty members within universities and National Institutes of Technology (KOSEN). The USA policies cater to a broader audience, including teachers, educational leaders, policymakers, researchers, and both innovators and providers of educational technology. Unlike China and Japan, which specifically address Generative AI, the USA encompasses a wider range of AI technologies, such as computer vision, robotics and Generative AI. Regarding the educational stages, China's policies do not specify a particular focus, Japan concentrates on higher education, and the USA targets primarily K-12 education levels (Table 1).

## Comparison of contents

### *Attitude towards Generative AI*

All four countries have a positive attitude toward Generative AI, each recognizing the transformative potential of AI in various sectors, including education, though their approaches and emphases differ. In the USA, the Department of Education views AI as a crucial tool to enhance educational methods, increase scale, and reduce costs, reflecting a pragmatic approach to leveraging technology in education. China, on the other hand, demonstrates a proactive and supportive stance toward generative AI, as evidenced by its policy to encourage innovation, using the term "encouragement" (鼓励) multiple times to underline its commitment. Japan acknowledges the benefits of AI in boosting economic and societal productivity and convenience but emphasizes a cautious approach by addressing potential risks related to AI reliability and misuse, underscoring the need for guidelines and responsive measures tailored to the educational context. Mongolia is collaborating with international organizations and foreign governments in launching initiatives to adopt AI and digitalization in education. For example, the government is collaborating with UNESCO in developing strategic documents while the Ministry of Education and Science is collaborating with the Eduten platform of Finland for secondary schools' mathematics education (MEDS, 2023). While all countries are eager to integrate AI technology,

China, Japan, and the USA, have been more cautious in encouraging innovation with responsible governance. Mongolia, on the other hand, prioritizes cyber security and digital education but the policies have not been clearly reflected in the education sector.

***Human-centered approach***

The USA and Japan prioritize the human center, however, China and Mongolia prioritize national security. The USA and Japan both emphasize a human-centered approach to the integration of AI in education, though they articulate their policies differently. The USA's Department of Education explicitly rejects the notion that AI could replace teachers, emphasizing AI as a tool to enhance learning while preserving human dignity and agency. This approach places humans at the center, ensuring that educators remain pivotal in instructional decisions and that AI supports rather than supplants human roles. Meanwhile, Japan's strategy, as outlined in its 'Human-Centered AI Social Principles' (2019), focuses on safeguarding human rights and enhancing human capabilities. It stresses the importance of using AI to support educational goals without infringing on constitutional rights and encourages the development of AI literacy to prevent misuse and over-dependence on technology. On the other hand, China, although not explicitly focusing on human-centered AI, prioritizes national security and the protection of public and legal interests, aligning AI development with national stability and control. Mongolia emphasizes SDGs and equity and access in education while aiming to boost its economy through digital advancement (Vision2050, n.d.).

***Teaching and learning***

In the context of teaching and learning, the USA and Japan illustrate it directly, however, China's approach is more implicit, oversight at the institutional level rather than detailing specific educational applications. The languages used in Mongolian documents describe a "rosy" picture of AI and digitalization in education, emphasizing the "innovation" and its benefits seldomly talking about concerns. The USA emphasizes a strategic integration of AI that supports educators and enhances teaching practices, focusing on selecting AI technologies that align with specific educational goals and support teacher involvement in instructional decisions. Japan advocates for the use of Generative AI as a tool to assist in proactive learning and administrative efficiency while maintaining strict guidelines to ensure Generative AI does not replace traditional learning methods or infringe on ethical standards. It encourages transparency in Generative AI use by students and stresses the importance of varied assessment methods to complement Generative AI use in education. Meanwhile, China's stance is less detailed regarding specific educational applications of Generative AI, with only general mentions of education in the context of norm-setting and administrative references. In Mongolia, facing a lack of teachers, particularly in rural areas, and an overload of teachers' work, the Minister of Education Mr. Enkh-Amgalan in April 2024 said the government is aiming to have "one AI teacher in one school" to increase access to education (MEDS, 2024).

***Diversity, equity, and inclusion (DEI)***

All four countries call on diversity, equity, and inclusion (DEI). The USA focuses on eliminating algorithmic discrimination, emphasizing the importance of addressing biases within AI algorithms to prevent unjust practices and ensure equitable learning opportunities, and human dignity, in education. Japan prioritizes the enhancement of life through AI, advocating for a society where diverse individuals can pursue happiness and sustainability, using AI to address societal disparities. Conversely, China targets the technical elements of AI development, implementing measures to prevent discrimination in algorithm design, training data, and service provision across various demographic factors. Mongolia mentions equity and access to education to deliver "high quality" education to students in rural areas through the use of AI and technology in general. However, Mongolia has no tangible programs implemented in practice. On the whole, the USA, Japan, and China demonstrate a commitment to integrating DEI principles in the deployment of AI in education and broader societal applications, each emphasizing different aspects from combating existing biases and enhancing human life to technical measures against discrimination tailored to their unique cultural and policy landscapes.

## Discussion

### National responses to Generative AI in higher education

The integration of AI in education across four countries reflects distinct national priorities, such as innovation, security, human-centric values, and economic advancement. The USA and Japan focus on enhancing educational methods and efficiency through a human-centered approach, using AI to support and augment human capabilities rather than replacing traditional educational roles. Conversely, China emphasizes innovation and aligns AI integration with national security, remaining less specific about its classroom applications and focusing more on overarching policy goals. Mongolia, while facing funding constraints that limit AI scalability, focuses its efforts on digital education for economic advancement and improving access in rural areas.

In addressing diversity, equity, and inclusion (DEI), each country attempts to adopt a tailored approach to mitigate potential biases and enhance educational outcomes through AI. The USA actively combats algorithmic biases to ensure fair AI usage, while Japan aims to improve the quality of life through cautious and reliable AI guidelines that protect educational integrity. China seeks to prevent technical biases to align with its security-focused AI policies, and Mongolia strives to improve educational access across diverse geographic areas, aiming to develop specific strategies to realize these goals. However, biases in AI are seldom discussed in the Mongolian discourses that we studied. Collectively, these strategies underscore the complex interplay between AI technology and educational policy, highlighting both the potential benefits and challenges of AI in global educational landscapes.

**Attention to the digital poverty gap**

Generative AI is built on extensive data sets and significant computing resources, along with continuous advancements in AI architectures and training techniques. These resources are predominantly accessible to major global tech firms and a select number of economies, primarily the United States, China, and Europe (Holmes & Miao, 2023). This implies that the capability to develop and manage generative AI is beyond the reach of most companies and countries, particularly those in the Global South. For example, Mongolian companies have developed AI tools and content for educational organizations to utilize, but the education sector faces a lack of funding to scale up and adopt the technologies. Since Generative AI is actively engaged in higher education, from 'data poverty' (Marwala, 2023) to digital poverty, Generative AI has the potential to worsen the digital divide between the Global North and South in higher education settings. However, in this study, in the educational AI policies of China, Japan, Mongolia, and the USA, a common issue is the lack of clear discussion and measures to address the digital divide. Therefore, policymakers at both national and international levels should be cognizant of the potential digital poverty gap and strive to enact policies that promote greater justice and fairness.

**Highlight the human-centered and diverse approach**

Generative AI based on the large language models, and the large language models like OpenAI's GPT series are predominantly trained on English language data (Kalyan, 2024) and embedded in Western culture (Karpouzis, 2024). Globally, as AI technology becomes more widespread in education, there is a risk of overlooking local cultural and linguistic characteristics. This could not only lead to the inapplicability of educational tools and content but also exacerbate the unequal distribution of educational resources, preventing students from non-dominant cultural and linguistic backgrounds from fully benefiting from the educational opportunities brought by AI.

To effectively navigate the challenges and capitalize on the advantages of generative AI in education, policymakers should balance the relationship between human agency and machine support, global context and local context. At first, applying Generative AI in higher education, human-centeredness should be prioritized (Holmes & Miao, 2023). Generative AI should complement, not replace, the role of teachers in education, emphasizing a human-centered approach that ensures ethical, safe, equitable, and meaningful integration. Concurrently, regulations should be developed to support the incorporation of local cultures and languages, enhancing the use of Generative AI within educational frameworks.

## Conclusions

AI is advancing rapidly, prompting societal shifts that necessitate a response through national policy. Given that policy development is a time-consuming process, policymakers and educational stakeholders must begin immediately to define the necessary requirements, disclosures, regulations, and frameworks. To unlock the potential of Generative AI in global higher education, future policies must not only focus on technological innovation but also consider infrastructural, socio-economic, and cultural factors to ensure the equitable and inclusive use of Generative AI. These efforts will help ensure a safe and beneficial future for all involved parties, particularly students and teachers. Admittedly, the limited sample size is one of the primary limitations of this study. It may not fully represent the diverse perspectives and variations that would be present in a larger sample. Despite the limitations, this study contributes to the existing literature by providing a comparative analysis of the attitudes and policies toward generative AI in higher education between China, Japan, Mongolia, and the USA, offering policymakers a broader perspective that integrates a global view ensures that Generative AI policies in higher education foster inclusion rather than exclusion.